# Plasmonic multiple exciton generation

Jiantao Kong[1], Xueyuan Wu[1], Xin Wang[2,3], Michael J. Naughton[1,2], and Krzysztof Kempa[1,4]

[1] *Department of Physics, Boston College, Chestnut Hill, Massachusetts 02467, USA*

[2] *International Academy of Optoelectronics at Zhaoqing, South China Normal University, Guangdong Province, China*

[3] *National Center for International Research on Green Optoelectronics, South China Normal University, Guangzhou, China*

[4] *Academy of Advanced optoelectronics, South China Normal University, Guangzhou, China*

**Abstract**

We show that bi-exciton formation can be highly efficient in a solar cell with the semiconductor absorber filled with an array of metallic nanoparticles having plasmonic resonance tuned to the semiconductor gap energy. This process can be viewed as plasmon-enhanced multiple exciton generation (PMEG), with the resulting cell efficiency exceeding the Shockley–Queisser limit. We demonstrate, that efficiency of the PMEG process, increases with decreasing of the semiconductor gap size, and illustrate that by considering in detail three systems with gradually decreasing gap size: GaAs, Si and Ge.

Electrons or holes in semiconductors, excited into the respective conduction and valence bands away from the thermal equilibrium distributions, are referred to as "hot". Effects of hot electrons have been studied and utilized for more than half a century in a variety of electronic devices, from Gunn diodes to integrated circuits [1-10]. In conventional solar cells, hot electrons rapidly and irreversibly lose their hot energy to phonons (heat), which leads to the Shockley-Queisser limit for single junction cell efficiency [11]. The amount of the energy lost to heat in a conventional device exceeds that harvested in the form of usable electricity. For example, commercially available, high efficiency crystalline silicon solar cells convert 20-25% of absorbed sunlight into electricity, but more than 30% into heat via hot electrons. Many concepts have been proposed to harvest or convert this hot electron energy into usable form, but none have been experimentally verified or demonstrated to date [11]. One of the seminal concepts proposed for so called third-generation solar photovoltaics (PV) involves harvesting the excess energy of these hot electrons before it is dissipated as heat [12], with theoretical efficiency limits of over 60%. This is posited to be achievable by first somehow eliminating the phonon scattering in the active region, and then extracting the hot electrons through narrow band energy filters at absorber-electrode contacts, assuring isentropic cooling. However, this is far from a trivial proposition, and no successful solar cell based on this idea has been developed. While early investigations found some evidence for hot electron injection into an electrolyte [13], and recently the hot electron contribution to the photo-voltage demonstrated [14], there remains limited experimental evidence of improved photovoltaic performance via hot electrons, despite many decades of research.

In another important scheme to recover the hot electron energy, it was envisioned that a single photon in a solar cell could generate two or more electron-hole pairs (physically-separated excitons), instead of a single pair. This is the multi-exciton generation (MEG) concept [15-17], known to be vanishingly small in bulk materials in the frequency range of interest to photovoltaics. It has been demonstrated in laser spectroscopic [16-17] and photocurrent [18] studies that, in semiconductor nanoparticles, it can become significant.

Recently, some of the present authors proposed a plasmonic 3rd generation PV scheme by providing an efficient energy-dissipation channel into plasmons in an adjacent or embedded plasmonic structure [19]. In this scheme, the hot electron free energy remains reversibly "protected" in a collective electronic degree of freedom. This hot electron plasmon protection (HELPP) mechanism, which relies on electron-plasmon scattering occurring on a time scale sufficiently smaller than phonon emission by either plasmons or hot electrons, was theoretically supported by a simple model calculation [19]. Here, we describe a way to combine the HELPP idea with MEG, a process which can be viewed as plasmon-enhanced multiple exciton generation (PMEG).

The MEG theory often breaks the process into two steps: first, an incoming photon excites a single exciton, with hot carriers participating; second, this exciton, before emitting phonons, decays into multiple excitons via Coulomb scattering [20]. Instead of employing Fermi's golden rule to estimate the decay rate of excitons (hot electrons and holes) to biexcitons, we calculate the hot electron scattering rate exactly, including the secondary excitons as a part of the single particle excitation continuum. The scattering rate of an electron in a semiconductor matrix, from a state $E_\mathbf{k}$ to states $E_\mathbf{k+q}$, due to single particle and collective (plasmon) excitations (with wave vectors *q*), is given in RPA by [21]

$$\gamma_{el-el} = \frac{2}{\hbar} \int \frac{d\mathbf{q}}{(2\pi)^3} V_q \left[ n_B\left(E_\mathbf{k} - E_\mathbf{k+q}\right) - n_F\left(-E_\mathbf{k+q} + \mu\right) \right] \text{Im}\left[ \varepsilon\left(q, \frac{E_\mathbf{k+q} - E_\mathbf{k}}{\hbar}\right)^{-1} \right] \qquad (1)$$

where $n_B$ and $n_F$ are the Bose-Einstein and Fermi-Dirac distribution functions, respectively, $\mu$ is the chemical potential, $\varepsilon(q,\omega)$ is the effective longitudinal dielectric function of the medium, and $V_q$ is the bare Coulomb interaction. Clearly, this calculation requires knowledge of the effective dielectric function of a given structure. In a simple, single Lorentzian approximation, the dielectric function can be written as [22]

$$\varepsilon(\omega) = \varepsilon_b + \frac{\omega_p^2}{\omega_r^2 - \omega(\omega + i\gamma)} \qquad (2)$$

which, for $\gamma \to 0^+$ and $\omega_r^2 \gg \omega_p^2$, when inserted into Eq. (1), leads to a simple formula [23]

$$\gamma_{el-el} \approx \frac{\sqrt{2E_k/m}}{2a^*} f\left(\frac{E_k}{\hbar\omega_r}\right) \Theta\left(\frac{E_k}{\hbar\omega_r} - 1\right) \qquad (3)$$

where the renormalized Bohr radius is $a^* = a_B \varepsilon_b^2 (\omega_r/\omega_p)^2$, and the auxiliary function $f(x) = \frac{2}{x}\ln\left(\sqrt{x} + \sqrt{x-1}\right)$ is slowly varying for $x > 1.5$. Eq. (3) can be used as guidance for more rigorous calculations/simulations, and it shows, as expected, that the scattering vanishes for $E_k < \hbar\omega_r$, and also that it increases rapidly with increasing plasmonic oscillator strength $\omega_p$.

Consider now a PV semiconductor absorber filled with a cubic array of simple spherical nanoparticles (nanospheres). We chose the period of the nanosphere lattice to be *a*, and the nanosphere diameter $D = a/3$, so that the projected area fraction remains unchanged as we change $a$. The normalized absorbance (ratio of the light absorbed to the normalized for all cases incoming flux), as simulated by employing FDTD code [24, 25] for crystalline GaAs semiconductor and silver nanoparticles, is shown in Fig. 1, for four values of $a$.

Fig. 1 shows, that the frequency of the plasmonic absorption increases with decreasing $a$, and saturates ~400 THz. This behavior reflects the well known dispersion relation of the surface plasmon, induced on the surface of the metallic sphere; changing sphere diameter changes an effective surface plasmon quasi-momentum according to the "whispering gallery" mode condition [14, 22] $q \approx 2/D$. The plasmonic absorption peak strengths rapidly increases, once the peak frequency enters the intersubband transition region above the gap energy of 1.4 eV (~340 THz). In this region, massive generation of interband transitions (excitons) by decaying hot electrons is also expected, and will be demonstrated below. The absorption spectrum for each value of $a$ is dominated by a single plasmonic resonance, and so one could use Eq. (2) as a simple model of the dielectric function, and then use Eq. (3) as a rough estimate of the scattering rate. For an accurate analysis we extract the effective dielectric function of the medium by the method described in detail in [26], and then use the exact Eq. (1) to obtain the scattering rate. The extracted single Lorentzian dielectric functions for $D = 67$ nm and 6.7 nm are shown in Fig. 2. The inset shows the corresponding scattering rates vs. hot electron energy. For the smaller spheres, intersubband transitions are possible (producing secondary excitons), and the scattering rates of hot electrons with energies 2.5 eV and more above the conduction band edge exceed 2 x $10^{13}$ s$^{-1}$. This is larger than the phonon cooling rate in GaAs of ~ 0.5x10$^{13}$ s$^{-1}$ [27]. This is the rate of cooling the hot electrons down to the bottom of the conduction band, which requires many electron-phonon scattering events; the energy of a single phonon is only ~ 36 meV, and so ~55 scattering events are needed to completely cool down a hot electron with energy 2 eV. The shaded area in the inset in Fig. 2, shows an estimated cooling rate. For larger spheres ($D = 67$ nm), with resonances below the energy gap, no secondary excitons are generated, only plasmons at a smaller rate.

The efficiency of this PMEG process diminishes with increasing the gap size; clearly only hot electrons with energy greater than the gap can generate secondary excitons. In fact, GaAs is not a suitable material for PMEG solar cells. The maximum value of the hot electron energy generated by the one-sun solar radiation (as measured from the top of the valence band) is about 3.4 eV [28], and so we estimate that in GaAs the hot electrons reach only up to about 3.4 eV – 1.4 eV = 2 eV into the conduction band. However, Fig. 2 shows that significant (exceeding the phonon scattering rate) plasmon generation occurs for hot electrons with energy > 2 eV, there is possibly only a tiny fraction sun-generated hot electrons which can generate secondary excitons. Nevertheless, GaAs is a good material to demonstrate the PMEG effect by using laser illumination.

Next, we investigate the crystalline Si. Employing exactly the same procedure as in the case of GaAs, we obtain the result shown in Fig. 3. The scattering rates are shown in the main part of Fig. 3 for two nanoparticle diameters D = 67 and 76 nm. In this case, we have the solar radiation induced hot electron band-width equal to 3.4 eV – 1.1 eV = 2.3 eV. For the larger diameter sphere, we obtain a significant scattering rate ($\sim 1.5 \times 10^{13}$ sec$^{-1}$) already for 1.3 eV, which exceeds that of the electron-phonon cooling rate ($< 10^{13}$ sec$^{-1}$). Thus, in this case a reasonably large portion of the hot electron distribution, of about 43%, is available for the PMEG recovery. Thus crystalline Si is a viable material for both, the PMEG demonstration, as well as for a PMEG solar cell.

Semiconductors with even smaller gaps, such as Ge (0.68 eV) or InAs (0.32 eV), should further improve efficiency of the PMEG process. As an example, we consider here Ge. Fig. 4 shows the electron-electron scattering rate, obtained by using the effective dielectric function shown in the inset, calculated for nanoparticles with $D = 33.3$ nm. The scattering rate has a maximum, representing the PMEG process at about 1.5 eV. Since in this case the range of hot electrons induced by a one-sun illumination is 3.4 eV – 0.7 eV = 2.7 eV (as measured from the bottom of the conduction band), a large fraction of hot electrons (more than 50%), with energies ranging from 1.3 eV to 2.7 eV can produce the secondary electrons. The electron-phonon scattering rate in Ge is $\sim 10^{14}$ sec$^{-1}$ [29], and the corresponding cooling rate (in view of the single phonon emission energy of ~20 meV [30]) is $\sim 10^{12}$ sec$^{-1}$, and therefore much lower than the electron-electron scattering rate. Thus we conclude, that Ge could be used as a practical platform for PMEG cells.

Finally, we comment on possible methods of developing arrays of NP inside active area. Wet chemistry processed semiconductors are the easiest, and the embedding can be achieved by simply mixing the NP with the semiconductor. Embedding NP into amorphous semiconductors processed by PECVD (a-Si and a-Ge) can be also obtained relatively easy by the layer-by-layer processing [31], or co-sputtering of a metal and semiconductor, followed by thermal processing [32]. Embedding plasmonic nanoparticles into crystalline semiconductors is much more challenging. Most promising are crystalline NP of silicides, which are plasmonic (metallic) with plasma energies in the 3 eV range [33], and so similar to Ag or Au. Most importantly silicides are lattice matched to Si, and so they can be epitaxially grown on Si [34], and vice versa [35]. Many of the silicide NP are also compatible with Ge, opening an avenue to PMEG solar cells. Another emerging technology is NP implantation, which allows deposition of NP growth seeds into semiconductors by ion implantation, and subsequent NP growth from those seeds during annealing, which restores crystalline structure [36].

In conclusion, we show that two-pair (bi-exciton) formation can be protected against phonon emission, and therefore be a likely event, if the semiconductor is filled with metallic nanoparticles having plasmonic resonance tuned to the semiconductor gap energy. The bi-exciton formation process then results from a rapid sequence of two events: (i) initial exciton generation by the incoming photon, and (ii) the second exciton generation by the plasmon-stimulated hot electron's decay. This process can be viewed as plasmon-enhanced multiple exciton generation (PMEG). The universality of this effect provides a new paradigm in the development of ultrahigh efficiency solar cells, beyond the Shockley–Queisser limit. We also demonstrate, that PMEG solar cells benefit from smaller gap semiconductors, and consider in detail three systems: large gap GaAs, intermediate gap $c$-Si and low gap Ge. While the first can be used only to demonstrate the PMEG process, the second and third could provide a possible platform for PMEG solar cells.

# References

[1] J. B. Gunn, "Microwave oscillation of current in III-V semiconductors", Solid State Commun. **1**, 88-91 (1963). doi: 10.1016/0038-1098(63)90041-3

[2] H. Kroemer, "Theory of the Gunn effect", Proc. IEEE **52**, 1736 (1964).

[3] D. Frohman-Bentchkowsky, "Memory behavior in a floating-gate avalanche-injection MOS (FAMOS) structure", Appl. Phys. Lett. **18**, 332-334 (1971). doi: 10.1063/1.1653685

[4] M. Heiblum, M. I. Nathan, D. C. Thomas, and C. M. Knoedler, "Direct observation of ballistic transport in GaAs", Phys. Rev. Lett. **55**, 2200-2203 (1985). doi: 10.1103/PhysRevLett.55.2200

[5] C. Rauch, G. Strasser, K. Unterrainer, W. Boxleitner, K. Kempa, and E. Gornik, "Ballistic electron spectroscopy of vertical biased superlattices", Physica E **2,** 282-286 (1998). doi: 10.1016/S1386-9477(98)00059-9

[6] A. Othonos, "Probing ultrafast carrier and phonon dynamics in semiconductors", J. Appl. Phys. **83**, 1789-1830 (1998). doi: 10.1063/1.367411

[7] J. R. Goldman and J. A. Prybyla, "Ultrafast dynamics of laser-excited electron distributions in silicon", Phys. Rev. Lett. **72**, 13641367 (1994). doi: 10.1103/PhysRevLett.72.1364

[8] S. M. Sze, "Microwave diodes", Chap. 9 in *High-Speed Semiconductor Devices*, Ed. by S. M. Sze (Wiley, New York 1990).

[9] K. M. Kramer and W. N. G. Hitchon, *Semiconductor Devices: A Simulation Approach* (Prentice Hall PTR, Upper Saddle River, New Jersey, 1997).

[10] J. A. Kash and J. C. Tsang, "Watching chips work: picosecond hot electron light emission from integrated circuits", J. Cryst. Growth **210**, 318-322 (2000). doi:10.1016/S0022-0248(99)00704-6

[11] M. A. Green, *Third Generation Photovoltaics: Advanced Solar Energy Conversion* (Springer, 2006).

[12] R. T. Ross and A. J. Nozik, "Efficiency of hot-carrier solar energy converters", J. Appl. Phys. **53**, 3813-3818 (1982). doi: 10.1063/1.331124

[13] J. A. Turner, A.J. Nozik, "Evidence for hot-electron injection across p-GaP/electrolyte junctions", Appl. Phys. Lett. **41,** 101-103 (1982). doi: 10.1063/1.93317

[14] K. Kempa, M. J. Naughton, Z. F. Ren, A. Herczynski, T. Kirkpatrick, J. Rybczynski, and Y. Gao, "Hot electron effect in nanoscopically thin photovoltaic junctions", Appl. Phys. Lett. **95**, 233121 (2009).

[15] M. C. Beard, J. M. Luther, O. E. Semonin, A. J. Nozik, "Third generation photovoltaics based on multiple exciton generation in quantum confined semiconductors", Acc. Chem. Res. **46**, 1252–1260 (2013). doi: 10.1021/ar3001958

[16] R. Schaller, V. Klimov, "High efficiency carrier multiplication in PbSe nanocrystals: Implications for Solar Energy Conversion", Phys. Rev. Lett. **92**, 186601 (2004).

[17] M. C. Beard, J. M. Luther, A. J. Nozik, "The promise and challenge of nanostructured solar cells", Nature Nano. **9**, 951–954 (2014). doi:10.1038/nnano.2014.292

[18] O. E. Semonin, J. M. Luther, S. Choi, H.-Y. Chen, J. Gao, A. J. Nozik, and M. C. Beard, Science **334**, 1530 (2011).

[19] J. Kong, A. H. Rose, C. Yang, J. M. Merlo, M. J. Burns, M. J. Naughton, K. Kempa, "A hot electron plasmon-protected solar cell", Opt. Exp. **23**, A1087-A1095 (2015). doi: 10.1364/OE.23.0A1087

[20] M. Voros, G. Galli and G. T. Zimanyi, "Colloidal nanoparticles for intermediate band solar cells", ACS Nano **9**, 6882-6890 (2015). doi: 10.1021/acsnano.5b00332

[21] G. D. Mahan, *Many-Particle Physics* (Plenum Press, New York 1981).

[22] Y. Wang, E. Plummer and K. Kempa, "Foundations of plasmonics," Adv. Phys. **60**, 799–898, (2011).

[23] K. Kempa, "Plasmonic protection of the electron energy", Phys. Stat. Solidi RRL **7**, 465 (2013).

[24] A. Taflove and S. C. Hagness, "Computational Electrodynamics: The Finite-Difference Time-Domain Method", Artech, Norwood MA (1995).

[25] CST Microwave Studio, the Computer Simulation Technology AG, https://www.cst.com/products/cstmws

[26] D. R. Smith, S. Schultz, P. Markoš, and C. M. Soukoulis, "Determination of effective permittivity and permeability of metamaterials from reflection and transmission coefficients", Phys. Rev. B **65**, 195104 (2002).

[27] M. Bernardi, D. Vigil-Fowler, C. S. Ong, J. B. Neaton, and S. G. Louie, "Ab initio study of hot electrons in GaAs", PNAS **112**, 5291-5296 (2015).

[28] M. Bernardi, D. Vigil-Fowler, J. Lischner, J. B. Neaton, and S. G. Louie, "Ab Initio Study of Hot Carriers in the First Picosecond after Sunlight Absorption in Silicon", PRL **112**, 257402 (2014).

[29] N. Tandon, J. D. Albrecht and L. R. Ram-Mohan, "Electron-phonon interaction and scattering in Si and Ge: Implications for phonon engineering" Journal of Applied Physics **118**, 045713 (2015).

[30] R. T. Payne, "Phonon Energies in Germanium from Phonon-Assisted Tunneling", Physical Review **139**, A570 (1965).

[31] Y. Zhang, B. Cai and B. Jia, *Nanomaterials* **6**, 95 (2016); doi:10.3390/nano6060095

[32] Sachan *et al., Nanomaterials and Energy* (2012), Volume **2** Issue NME1.

[33] R. Soref, R. E. Peale, and W. Buchwald, "Longwave plasmonics on doped silicon and silicides", *OPTICS EXPRESS*, 28 April 2008 / Vol. 16, No. 9 / pp. 6507.

[34] J. J. Chen, "Metal silicides: An Integral Part of Microelectronics", *JOM* (2005), vol. 57, no 9. pp = 24 -31.

[35] J.- Y. Veuillen, "Growth of silicon thin films on erbium silicide by solid phase epitaxy", Journal of Applied Physics **75**, 223 (1994); https://doi.org/10.1063/1.355887

[36] H. W. Seo, et al., "Formation of silver nanoparticles in silicon by metal vapor vacuum arc ion implantation", *Nuclear Instruments and Methods in Physics Research* B **292** (2012) 50–54.

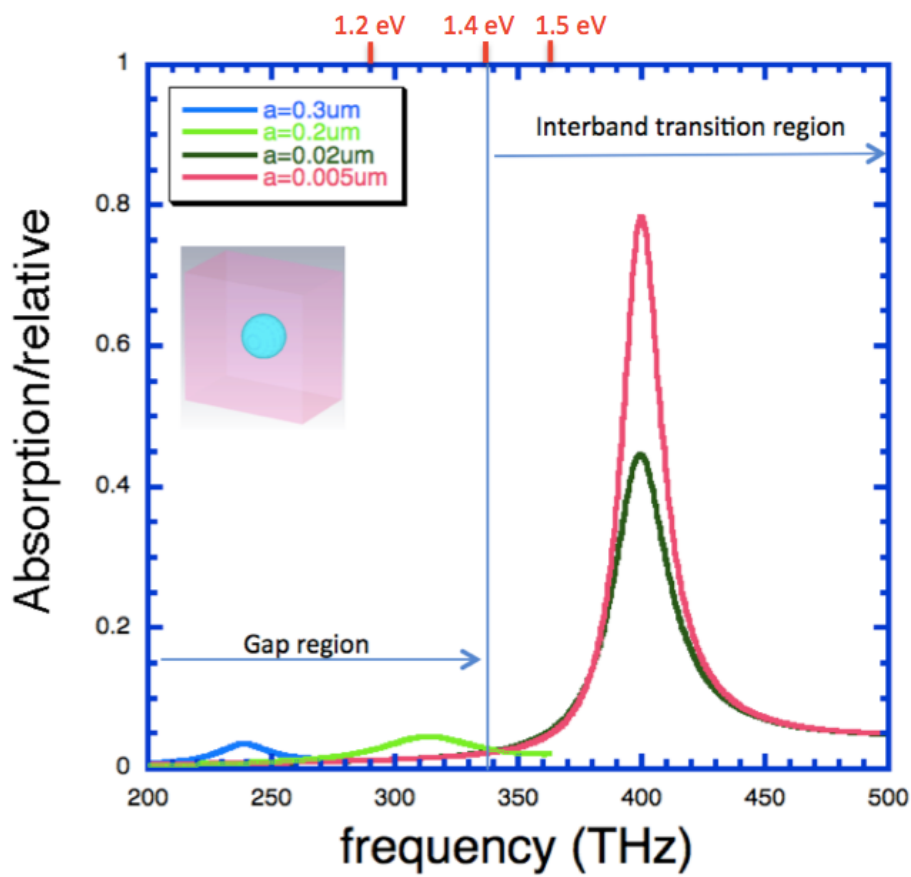

**Fig. 1.** Normalized absorbance spectra of the GaAs absorber filled with a cubic array of silver nanospheres (each with diameter $D = a/3$)

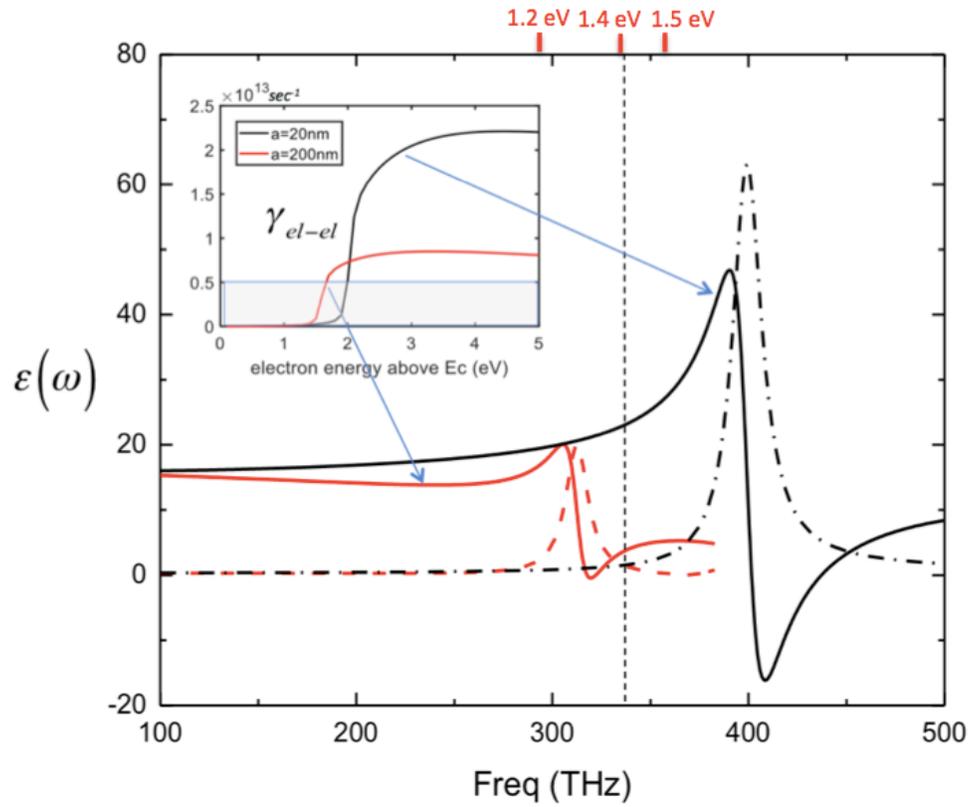

**Fig. 2** Extracted effective dielectric function of the GaAs absorber filled with a cubic array of silver nanospheres (each with diameter $D = a/3$) for two nanosphere sizes $D = 6.7$ nm (black), and $a = 67$ nm (red). The inset shows the corresponding electron- electron scattering rates. The shaded area represents the rates of electron-phonon scattering processes.

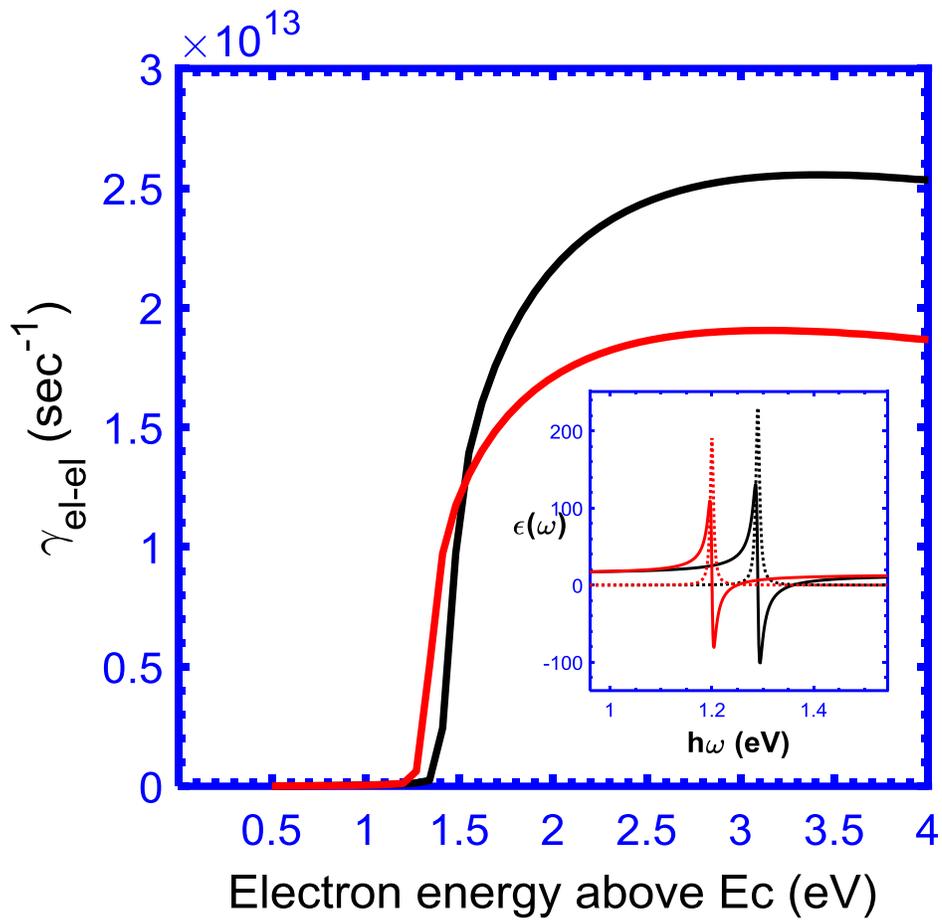

**Fig. 3** The calculated electron-electron scattering rates for a crystalline silicon absorber filled with a cubic array of silver nanospheres (each with diameter $D = a/3$). 200 nm (black curve) and $a$ =230 nm (red curve). The inset shows the corresponding extracted effective dielectric function, used to obtain the scattering rates.

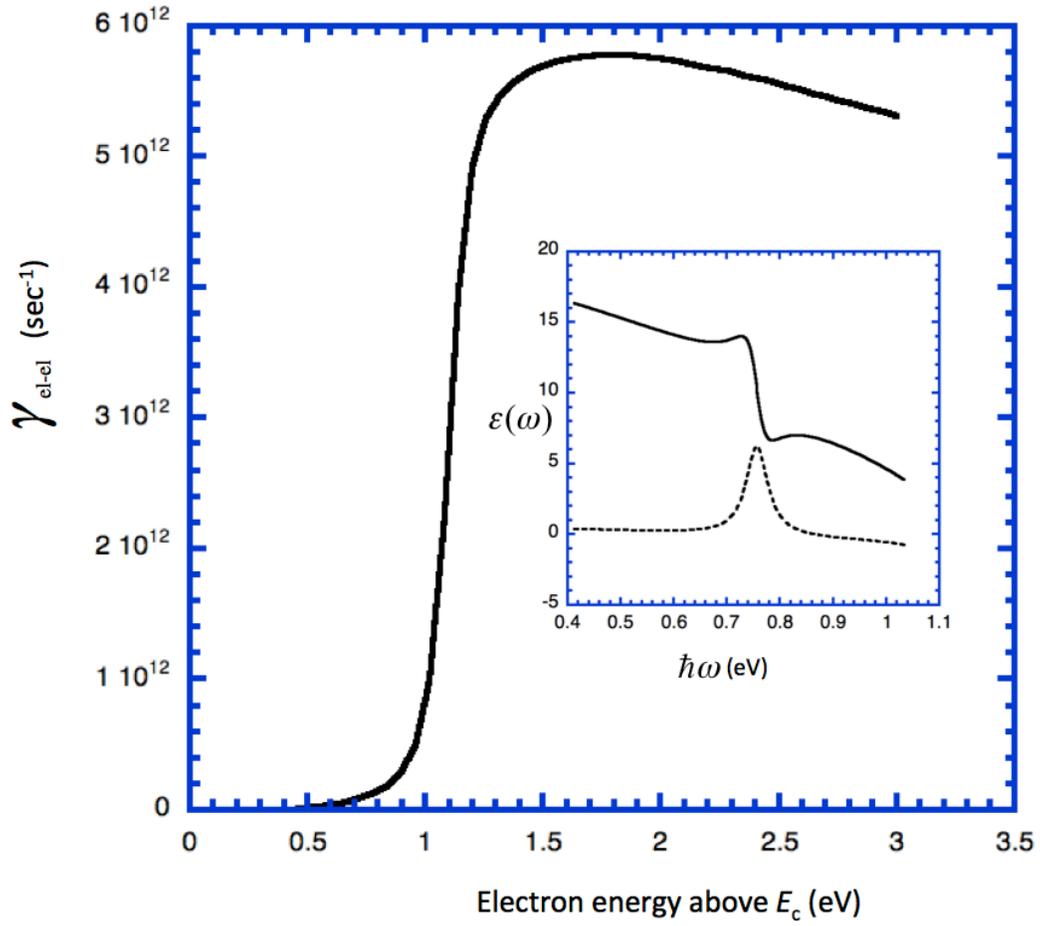

**Fig. 4** The calculated electron- electron scattering rates for a Ge absorber filled with a cubic array of silver nanospheres (each with diameter $D = a/3 = 33.3$ nm). The inset shows the corresponding extracted effective dielectric function, used to obtain the scattering rates.